\documentclass[aps,pre,showpacs,twocolumn]{revtex4-1}
\usepackage{amsmath} % math extension
\usepackage{amssymb} % symbols
\usepackage{graphicx}% include figure files
\usepackage{color}

\begin{document}

\title{Stress-stress fluctuation formula for elastic constants in the NPT ensemble}
\author{Dominik Lips}
\author{Philipp Maass}
\email{maass@uos.de}
\affiliation{Fachbereich Physik, Universit\"at Osnabr\"uck,
             Barbarastra{\ss}e 7, 49076 Osnabr\"uck, Germany}

%\date{11 September 2017, Last revised 9 May 2018}

\begin{abstract}

Several fluctuation formulas are available for calculating elastic constants from equilibrium correlation functions 
in computer simulations, but the ones available for simulations at constant pressure exhibit slow
convergence properties and cannot be used for the determination of local elastic constants. 
To overcome these drawbacks, we derive a stress-stress fluctuation formula in the $NPT$ ensemble based on
known expressions in the $NVT$ ensemble. We validate the formula in the $NPT$ ensemble by 
calculating elastic constants for the simple nearest-neighbor Lennard-Jones crystal and by 
comparing the results with those obtained in the $NVT$ ensemble. For both local and bulk elastic constants 
we find an excellent agreement between the simulated data in the two ensembles. To demonstrate the
usefulness of the new formula, we apply it to determine the elastic constants of a simulated lipid bilayer.
\end{abstract}

%05.10.-a	 Computational methods in statistical physics and nonlinear dynamics (see also 02.70.-c in mathematical methods in physics)
%62.20.de Elastic moduli
%62.20.dq Other elastic constants
%81.40.Jj	Elasticity and anelasticity, stress-strain relations

\maketitle

\section{Introduction}
The possibility to determine elastic properties of solid and soft materials from computer simulations
\cite{Ray:1988, Barrat:2006}
has significantly advanced our understanding of how non-crystalline materials react under loading. This 
includes strongly inhomogeneous systems such as glasses \cite{MizunoMossaBarrat2013}, polymer fibers \cite{Meier1993}
and granular materials \cite{Aubouy2003}. Particularly useful is the access to local elastic constants and 
their spatial variation, which, for example, can provide important information on the change of elastic 
properties close to interfaces, grain boundaries, and surfaces \cite{VanWorkumDePablo2003, Kluge1990}.
For obtaining the elastic constants one can directly 
simulate stress-strain curves by applying small deformations (or stresses) 
\cite{Ray:1988, Griebel/Hamaekers:2004}, or
analyze equilibrium correlation functions by resorting to the fluctuation-dissipation theorem
\cite{Ray:1988, Parrinello/Rahman:1982}, or
introduce phenomenological expressions of free energies with parameters, which 
can be determined in simulations and related to the constants of the free energy expansion with 
respect to strain \cite{West/etal:2009, Neder/etal:2010}.

Most straightforward procedures for calculating elastic constants are
the evaluations of proper equilibrium correlation functions in fluctuation formulas (FF).
Various approaches with different levels of complexity have been established to this end.
Among them, the FF applicable in constant pressure simulations neither exhibit good 
convergence properties upon averaging nor do they allow one to determine \textit{local}
elastic constants, which are important for the characterization of inhomogeneous materials
\cite{Barrat:2006, Ganguly/etal:2013}.
A FF with good convergence properties and access to local elastic constants has been
derived for the $NVT$ ensemble \cite{Gusev/etal:1996, Lutsko:1988}, 
but no corresponding FF is yet available  for the $NPT$ ensemble.
The goal of this work is to fill this gap.

After given an overview of the existing methods to calculate elastic constants from FF in 
Sec.~\ref{sec:nvt}, we derive the new FF in the $NPT$ ensemble in Sec.~\ref{sec:npt}. In 
Sec.~\ref{sec:nnlj-application} we then validate the FF by calculating the elastic constants of the
nearest-neighbor Lennard-Jones fcc solid and comparing results with those reported earlier in the
literature. We further show that, to obtain well-converged values for the elastic constants, one needs to
perform averages over a comparable number of
equilibrated particle configurations in the $NPT$ and $NVT$ ensemble.
We demonstrate the usefulness of the new FF by determining the elastic constants of a simulated 
lipid bilayer based on the model developed in \cite{Lenz/Schmid:2007} in Sec.~\ref{sec:lipid-application}.
In this model, the bilayer is stabilized by a surrounding gas of solvent beads, reflecting the pressure 
exerted by an aqueous environment. Thus, the access to local elastic constants allows us to
selectively extract the elastic properties of the lipid bilayer, without the need to modify the 
most convenient simulation procedure of such a system at constant pressure.
As pointed out in the concluding Sec.~\ref{sec:conclusion}, this possibility will be one of 
the main advantages of the new FF.

\section{Fluctuation formulas for elastic constants}
\label{sec:nvt}
For calculating elastic constants in molecular dynamics simulations,
a special molecular dynamics ensemble with a 
fixed external stress tensor $\tau$, the $N\tau T$ ensemble ($N$: number of particles, $T$: temperature)
was introduced \cite{ParrinelloRahman1981}.
In this ensemble the shape of the simulation cell, and accordingly
the instantaneous strain tensor $\epsilon$, is allowed to 
fluctuate. In general, the pressure is given by the trace of $\tau$,
and the special choice $\tau_{\alpha\beta}=-P\delta_{\alpha\beta}$ corresponds 
to the $NPT$ ensemble.
The geometry of the rectilinear simulation cell is described
by a scaling matrix $h$, whose columns 
are the vectors along the edges of the cell. The first order bulk elastic constants 
$C_{\alpha \beta \mu \nu}$ can be calculated from 
the strain-strain FF \cite{Parrinello/Rahman:1982}
\begin{equation}
C_{\alpha \beta \mu \nu}^{-1} = \frac{k_{\rm\scriptscriptstyle B} T}{\langle V \rangle_\tau} 
\langle \epsilon_{\alpha \beta} \epsilon_{\mu \nu} \rangle_\tau\,,
\label{eq:strain-strain-ff}
\end{equation}
where $k_{\rm\scriptscriptstyle B}$ is the Boltzmann constant, $V$ the instantaneous volume, and 
$\langle\ldots\rangle_\tau$ 
denotes the average in the $N\tau T$ ensemble; $C_{\alpha \beta \mu \nu}^{-1}$ is the inverse of
$C_{\alpha\beta\mu\nu}$ in the sense that $C_{\alpha \beta \mu\nu}^{-1}C_{\mu\nu\gamma\eta}=\delta_{\alpha\gamma}\delta_{\beta\eta}$.
The strain tensor $\epsilon$ is given by the scaling matrix $h$
in the instantaneous geometry
and a reference geometry characterized by $h_0$, which corresponds to the mean
of the scaling matrix under a given external stress $\tau$. Specifically
$\epsilon=(h_0^\text{-1T}h^\text{T}h\,h_0^{-1}-1)/2$. 
The approach only requires a simulation in the $N\tau T$ ensemble to calculate the full tensor of isothermal elastic constants and no evaluation of specific particle interactions is necessary.
Information about the specific system under consideration enters indirectly through the phase space measure of the ensemble average.
Unfortunately, application of the strain-strain FF
results in poor convergence when averaging the strain-strain fluctuations.

To address this problem, Gusev \textit{et al.} introduced the stress-strain FF in
the $N \tau T$ ensemble \cite{Gusev/etal:1996}, 
\begin{equation}
C_{\alpha \beta \mu \nu} = 
\langle \epsilon_{\alpha \beta} \hat{t}_{\lambda \gamma} \rangle_\tau
\langle \epsilon_{\lambda \gamma} \epsilon_{\mu \nu} \rangle_\tau^{-1},
\label{eq:stress-strain-ff}
\end{equation}
where $\hat{t}_{\alpha\beta}$ is the tension operator whose average
$\langle\hat{t}_{\alpha\beta}\rangle_\tau$ gives 
the thermodynamic tensions (or second Piola-Kirchhoff stress). Here and in the following
we denote phase space operators, which are functions of the particle positions and momenta, 
by a caret. The tension operator $\hat{t}$ 
is related to the (Cauchy) stress operator $\hat\Sigma$ by \cite{Ray:1988}
\begin{equation}
\hat\Sigma =\frac{V_0}{V}\,hh_0^{-1}\hat t\, h_0^{-1\mathrm{T}}h^\mathrm{T}\,,
\label{eq:tension-stress}
\end{equation}
where $V=\det h$ and $V_0=\det h_0$ is the volume of the reference
geometry. At low temperatures
this approach offers improved convergence properties \cite{Gusev/etal:1996}. 
When using Eq.~(\ref{eq:stress-strain-ff}),
the specific form of interactions in the system enters directly via $\hat\Sigma$ 
and in turn  $\hat t$  in Eq.~(\ref{eq:tension-stress}). 
The stress operator depends on
the first derivatives of the potential energy (i.~e., the forces) and for pairwise interactions is
given by
\begin{equation}
\hat\Sigma_{\alpha \beta} = -\frac{1}{V} 
\left[
\sum_i \frac{p_{i, \alpha} p_{i, \beta}}{m_i}
-\sum_{i<j} \frac{\partial\hat U}{\partial r_{ij}}\frac{x_{ij, \alpha} x_{ij, \beta}}{r_{ij}}
\right] \,,
\label{eq:definition-bulk-stress-tensor}
\end{equation}
where $\vec r_i$ and $\vec p_i$ are, respectively, the positions and momenta of the $N$ particles, $\hat U=(1/2)\sum_{i,j}^N \hat u(\vec r_{ij})$ is the  potential energy with $\vec r_{ij}=(\vec r_i-\vec r_j)=(x_{ij,1},x_{ij,2},x_{ij,3})$, and $r_{ij}=|\vec{r}_{ij}|$.
In molecular dynamics simulations, the stress-strain FF increases the computational costs only insignificantly, since the forces are calculated anyway. 

In the $NVT$ ensemble,
the determination of the first order elastic constants can be achieved with the stress-stress FF \cite{Ray/Rahman:1984, Lutsko:1989}. 
It requires more computational effort, 
because second-order derivatives of the potential need to be evaluated.
For truncated forces, these second-order derivatives lead to 
$\delta$-function contributions that must be properly dealt with
\cite{Xu/etal:2012}. The advantage of the stress-stress FF is that the
numerical averaging of the respective correlation functions
can converge by orders of magnitude more rapidly compared to 
the strain-strain and stress-strain FF. 
Furthermore, 
Lutsko showed \cite{Lutsko:1988}
that it is possible to combine this method with Irving and Kirkwood's definition of a local stress tensor
\cite{IrvingKirkwood1950}, to obtain a local form of the stress-stress FF:
\begin{subequations}
\label{eq:stress-stress-ff-nvt}
\begin{align}
C_{\alpha \beta \mu \nu}(\vec{r}) &= 
C^\textrm{K}_{\alpha \beta \mu \nu}(\vec{r}) + 
C^\textrm{B}_{\alpha \beta \mu \nu}(\vec{r}) - 
C^\textrm{N}_{\alpha \beta \mu \nu}(\vec{r}), 
\label{eq:stress-stress-ff-nvt-a}\\
C^\textrm{K}_{\alpha \beta \mu \nu}(\vec{r}) 
&= 
2 \langle \hat\rho(\vec{r}) \rangle_{\scriptscriptstyle V} k_{\rm\scriptscriptstyle B} T 
(\delta_{\alpha\mu} \delta_{\beta\nu} + \delta_{\alpha\nu} \delta_{\beta\mu}), 
\label{eq:stress-stress-ff-nvt-b}\\
C^\textrm{B}_{\alpha \beta \mu \nu}(\vec{r}) 
&= 
\bigg\langle 
\sum_{i<j} \hat B^{(ij)}_{\alpha \beta \mu \nu} \, g(\vec{r};\vec{r}_{i}, \vec{r}_{j})
\bigg\rangle_{\scriptscriptstyle V}, 
\label{eq:stress-stress-ff-nvt-c}\\
C^\textrm{N}_{\alpha \beta \mu \nu}(\vec{r}) &= \frac{V}{k_{\rm\scriptscriptstyle B} T}
\Big[\langle \hat\sigma_{\alpha \beta}(\vec{r})\, \hat\Sigma_{\mu \nu} \rangle_{\scriptscriptstyle V} 
- \langle \hat\sigma_{\alpha \beta}(\vec{r}) \rangle_{\scriptscriptstyle V} \langle \hat\Sigma_{\mu \nu} \rangle_{\scriptscriptstyle V}\Big] ,
\label{eq:stress-stress-ff-nvt-d}	
\end{align}
\end{subequations}
where $\langle\ldots\rangle_{\scriptscriptstyle V}$ denotes the equilibrium average in the $NVT$ ensemble;
$C^\textrm{K}$ is the ideal gas contribution, $C^\textrm{B}$ is the Born-term describing the response to 
affine deformations, and the non-affine term $C^\textrm{N}$ accounts for internal relaxation. 
In Eq.~(\ref{eq:stress-stress-ff-nvt-b}), the density function is
\begin{equation}
\hat\rho(\vec{r})=\sum_i \delta(\vec{r}-\vec r_i)\,.
\label{eq:rhofunction}
\end{equation}
For pairwise interactions, the Born functions $\hat B^{(ij)}$ 
in Eq.~(\ref{eq:stress-stress-ff-nvt-c}) are
\begin{equation}
\hat B^{(ij)}_{\alpha \beta \mu \nu} = 
\left( \frac{\partial^2\hat U}{\partial r^2_{ij}} 
- \frac{1}{r_{ij}} \frac{\partial\hat U}{\partial r_{ij}}\right)
\frac{r_{ij, \alpha} r_{ij, \beta} r_{ij, \mu} r_{ij, \nu}}{r^3_{ij}}\,,
\label{eq:bornfunction}
\end{equation}
and the local stress operator in Eq.~(\ref{eq:stress-stress-ff-nvt-d}) is given by
\begin{align}
\hat\sigma_{\alpha \beta}(\vec{r}) = 
&- \sum_i \frac{p_{i, \alpha} p_{i, \beta}}{m_i} \delta(\vec{r} - \vec{r}_i) \nonumber\\
&{}+ \sum_{i<j} 
\frac{\partial\hat U}{\partial r_{ij}}
\frac{x_{ij, \alpha} x_{ij, \beta}}{r_{ij}}\,
g(\vec{r};\vec{r}_{i}, \vec{r}_j)\,.
\label{eq:definition-local-stress-tensor}
\end{align}
Here,  $g(\vec{r};\vec{r}_{i}, \vec{r}_j)$ is a weighting function,
which corresponds to a Dirac $\delta$-function of $\vec{r}$ 
with support on the line segment
joining the points $\vec{r}_i$ and $\vec{r}_j$ 
divided by the distance $r_{ij}$,
\begin{equation}
g(\vec{r};\vec{r}_{i}, \vec{r}_j)=
\frac{1}{r_{ij}}
\int_0^1\mathrm{d}\lambda\,\delta\big(\vec{r}-(1-\lambda)\vec{r}_i-\lambda\vec{r}_j\big)\,.
\end{equation}
The expressions in Eqs.~(\ref{eq:stress-stress-ff-nvt-a})-(\ref{eq:stress-stress-ff-nvt-d}) and (\ref{eq:definition-local-stress-tensor}) correspond to microscopic (operator-like) continuum fields, from which
by a local spatial averaging (coarse-graining) smooth continuum fields are obtained. In practice, the system is partitioned into a grid of cells and the elastic constants are calculated for each cell \cite{MizunoMossaBarrat2013, WorkumDePablo2004}.
Upon spatial averaging over the whole volume $V$, one obtains 
from Eq.~(\ref{eq:definition-local-stress-tensor}) the bulk stress tensor $\hat\Sigma$
and from Eq.~(\ref{eq:stress-stress-ff-nvt}) 
the bulk elastic constants. This amounts to replace $\langle \hat\rho(\vec{r}) \rangle_{\scriptscriptstyle V}$ by the bulk density $N/V$ in 
Eq.~(\ref{eq:stress-stress-ff-nvt-b}), to set $g=1$ in Eq.~(\ref{eq:stress-stress-ff-nvt-c}), and to replace $\hat\sigma$ by $\hat\Sigma$ 
in Eq.~(\ref{eq:stress-stress-ff-nvt-d}).

\section{Local elastic constants in the NPT ensemble}
\label{sec:npt}
Local elastic constants at a given pressure can be obtained if one first performs a simulation in 
the $NPT$ ensemble and determines the bulk density, and thereafter adjusts the volume
in the $NVT$ ensemble to that density \cite{Kanigel/etal:2001, Yoshimoto/etal:2005, Gao2006, Jun2007}.
Here we are interested in enabling a direct determination of local elastic constants in the $NPT$ ensemble.
To this end we will in the following
transform the stress-stress FF in the $NVT$ to the $NPT$ ensemble. A similar approach was given in 
\cite{Wittmer/etal:2013a, Wittmer/etal:2013b, Wittmer/etal:2013c}.

Let us first note that the average yielding the Born term in Eq.~(\ref{eq:stress-stress-ff-nvt-c}) is
independent of the chosen ensemble in the thermodynamic limit
as long as the interaction potential between the particles is sufficiently short-ranged \footnote{The pair potential
should decrease faster than $1/r^3$ with the particle distance $r$, or, for slower decaying pair potentials, 
screening effects should lead to an effective decay faster than $1/r^3$.}. This is because
the Born function in Eq.~(\ref{eq:bornfunction})
can, for sufficiently short-range interactions, be viewed as a sum over independent particle contributions with the
$g(\ldots)$-function scaling with $1/V$. The same reasoning applies to the density function in Eq.~(\ref{eq:rhofunction}),
yielding, in the thermodynamic limit, the ensemble-independent local density
in the kinetic term in Eq.~(\ref{eq:stress-stress-ff-nvt-b}).

However, correlations of phase space functions, like 
they occur in the non-affine term in Eq.~(\ref{eq:stress-stress-ff-nvt-d}),
are in general not equal for different ensembles in
the thermodynamic limit.
In Ref.~\cite{LebowitzPercusVerlet1967}, a formalism to relate correlations in different ensembles was
worked out. 
Applying this formalism to the covariance of the local 
and bulk stresses in Eq.~(\ref{eq:stress-stress-ff-nvt-d})
yields
\begin{align}
\big\langle\Delta \hat{\sigma}_{\alpha \beta}(\vec{r}) &
\Delta \hat{\Sigma}_{\mu \nu} \big\rangle_{\scriptscriptstyle V}
= 
\big\langle 
	\Delta \hat{\sigma}_{\alpha \beta} (\vec{r})
	\Delta \hat{ \Sigma}_{\mu \nu} 
\big\rangle_{\scriptscriptstyle P} \nonumber \\
&\hspace{-2em}{}+ 
k_{\rm \scriptscriptstyle B} T
\left( 
	\frac{\partial\langle V\rangle_{\scriptscriptstyle P}}{\partial P} 
\right)^{-1}
\left( 
	\frac{\partial}{\partial P} 
	\big\langle 
		\hat{\sigma}_{\alpha \beta}(\vec{r})
	\big\rangle_{\scriptscriptstyle P} 
\right) 
\left( 
	\frac{\partial}{\partial P} 
	\big\langle 
		\hat{ \Sigma}_{\mu \nu} 
	\big\rangle_{\scriptscriptstyle P} 
\right) \nonumber \\
&\hspace{-2em}{}+ \mathcal{O}(1/N) ,
\label{eq:covariance-in-different-ensembles}
\end{align}
where $\Delta X = X - \langle X \rangle$ denotes the deviation of the quantity 
$X$ from its average,
and $\langle\ldots\rangle_{\scriptscriptstyle P}$ is the average in the $NPT$ ensemble 
\footnote{This formula has the following meaning: In the $NVT$ ensemble the average pressure has the 
value $P$, and it is this pressure
that defines the $NPT$ ensemble for evaluating the average. The partial derivatives with respect to
$P$ refer to changes of the corresponding quantities
in the $NPT$ ensemble with $P$, and need to be evaluated at
the pressure $P$.}.

The three derivates in the second line of Eq.~(\ref{eq:covariance-in-different-ensembles})
are readily obtained from the fluctuation dissipation theorem,
\begin{subequations}
\begin{align}
k_{\rm\scriptscriptstyle B} T\, \frac{\partial
\big\langle 
\hat{\sigma}_{\alpha \beta} (\vec{r})
\big\rangle_{\scriptscriptstyle P} 
}{\partial P} &=
-\langle \Delta \hat{\sigma}_{\alpha \beta}(\vec{r}) \Delta V \rangle_{\scriptscriptstyle P} \,,
\label{eq:local-stress-pressure-derivative}\\
k_{\rm\scriptscriptstyle B} T\, \frac{\partial
\big\langle 
\hat{\Sigma}_{\alpha \beta}
\big\rangle_{\scriptscriptstyle P}}{\partial P} &=
-\langle \Delta \hat{\Sigma}_{\alpha \beta} \Delta V \rangle_{\scriptscriptstyle P}
=-k_{\rm\scriptscriptstyle B} T\delta_{\alpha\beta}\,,
\label{eq:global-stress-pressure-derivative}
\end{align}
\end{subequations}
where Eq.~(\ref{eq:global-stress-pressure-derivative}) also follows 
by resorting to the expression
$\langle \hat{ \Sigma}_{\alpha\beta} \rangle_{\scriptscriptstyle P} = -P\delta_{\alpha\beta}$
for hydrostatic pressure. Putting everything together,
 we obtain the following stress-stress FF for the local elastic constants
in the $NPT$ ensemble:
\begin{subequations}
\label{eq:stress-stress-ff-npt}
\begin{align}
C_{\alpha \beta \mu \nu}(\vec{r}) &= 
C^\textrm{K}_{\alpha \beta \mu \nu}(\vec{r}) + 
C^\textrm{B}_{\alpha \beta \mu \nu}(\vec{r}) - 
C^\textrm{N}_{\alpha \beta \mu \nu}(\vec{r}), 
\label{eq:stress-stress-ff-npt-a}\\
C^\textrm{K}_{\alpha \beta \mu \nu}(\vec{r}) 
&= 
2 \langle \hat\rho(\vec{r}) \rangle_{\scriptscriptstyle P} k_{\rm\scriptscriptstyle B} T 
(\delta_{\alpha\mu} \delta_{\beta\nu} + \delta_{\alpha\nu} \delta_{\beta\mu}), 
\label{eq:stress-stress-ff-npt-b}\\
C^\textrm{B}_{\alpha \beta \mu \nu}(\vec{r}) 
&= 
\bigg\langle 
\sum_{i<j} \hat B^{(ij)}_{\alpha \beta \mu \nu} \, g(\vec{r};\vec{r}_{i}, \vec{r}_{j})
\bigg\rangle_{\scriptscriptstyle P}, 
\label{eq:stress-stress-ff-npt-c}\\
C^\textrm{N}_{\alpha \beta \mu \nu}(\vec{r})&= 
\frac{\langle V \rangle_{\scriptscriptstyle P}}{k_{\rm\scriptscriptstyle B} T}
\big\langle \Delta \hat{\sigma}_{\alpha \beta}(\vec{r}) \Delta \hat{\Sigma}_{\mu \nu} 
\big\rangle_{\scriptscriptstyle P}
\nonumber \\
&\phantom{=}{}
-\frac{K}{k_{\rm\scriptscriptstyle B}T}
\big\langle 
\Delta \hat{\sigma}_{\alpha \beta}(\vec{r}) \Delta V 
\big\rangle_{\scriptscriptstyle P}\,
\delta_{\mu \nu} .
\label{eq:stress-stress-ff-npt-d}	
\end{align}
\end{subequations}
In Eq.~(\ref{eq:stress-stress-ff-npt-d}), 
we have replaced the volume $V$ from Eq.~(\ref{eq:stress-stress-ff-nvt-d}) by the corresponding
average $\langle V\rangle_{\scriptscriptstyle P}$ in the $NPT$ ensemble (in the thermodynamic limit, 
$V=\langle V\rangle_{\scriptscriptstyle P}$),
and inserted the isothermal bulk modulus 
\begin{equation}
K=-V\frac{\partial P}{\partial V}=
k_{\rm\scriptscriptstyle B}T
\frac{\langle V\rangle_{\scriptscriptstyle P}}{\langle \Delta V^2\rangle_{\scriptscriptstyle P}}\,.
\end{equation}
All quantities in Eqs.~(\ref{eq:stress-stress-ff-npt})
can be sampled in a single simulation run in the $NPT$ ensemble.

Finally, we point out that the elements  $C_{\alpha \beta \mu \nu}$
give the coefficients of the second order term in an expansion
of the free energy with respect to strain, while
elastic constants $\tilde C_{\alpha \beta \mu \nu}$ 
in the sense of Hooke's law relate the stress $\sigma$ to the
linearized strain $\tilde\epsilon=(h_0^\text{T\,-1}h^\text{T}+hh_0^{-1})/2-1$ valid
for small deformation gradients.  
In an initial stress-free configuration
(vanishing first oder term of the free energy expansion),
the two tensors agree, $\tilde C=C$. However,
in a stressed reference configuration,
an extra linear term has to be taken into account in the free energy expansion.
This implies that the two tensors are no longer equal, but are related via \cite{Barron1965} 
\begin{align}
\tilde C_{\alpha \beta \mu \nu} (\vec{r}) &=C_{\alpha \beta \mu \nu} (\vec{r}) +
\frac{1}{2} 
\Big(
\langle \hat{\sigma}_{\alpha \mu} (\vec{r}) \rangle \delta_{\beta \nu}
+\langle \hat{\sigma}_{\alpha \nu} (\vec{r}) \rangle \delta_{\beta \mu}\nonumber\\
&\hspace{-3em}+\langle \hat{\sigma}_{\beta \mu} (\vec{r}) \rangle \delta_{\alpha \nu}
+\langle \hat{\sigma}_{\beta \nu} (\vec{r}) \rangle \delta_{\alpha \mu}-2 \langle \hat{\sigma}_{\alpha \beta} (\vec{r}) \rangle \delta_{\mu \nu}
\Big) .\end{align}
In particular, the additional term on the right hand side has to be taken into account for
simulations under a finite pressure.

The expansion coefficient $C_{\alpha \beta \mu \nu}$ 
and the elastic constants $\tilde C_{\alpha \beta \mu \nu}$ both in their local form and bulk form 
exhibit the symmetries $C_{\alpha \beta \mu \nu}=C_{\beta\alpha\mu \nu}=C_{\alpha \beta\nu\mu }$.
These symmetries reduce the $3^4=81$ coefficients to 36 independent ones that in the usual
Voigt notation \cite{Voigt1910} are represented by a $6\times6$ matrix, which generally can be asymmetric.
The bulk expansion coefficients $C_{\alpha \beta \mu \nu}$ 
exhibit the additional symmetry
$C_{\alpha \beta \mu \nu}=C_{\mu \nu\alpha \beta }$,
which implies that
the corresponding matrix in Voigt notation becomes symmetric.
The elastic constants $\tilde C_{\alpha \beta \mu \nu}$ in general do not have this additional symmetry,
unless for hydrostatic pressure, where $\langle\hat\Sigma_{\alpha\beta}\rangle=-P\delta_{\alpha\beta}$
(including stress-free reference configurations). 
Additional symmetries are reflecting symmetries of the material structure.

Let us finally note that care should be taken when integrating out the momenta in the non-affine term
$C^\textrm{N}_{\alpha \beta \mu \nu}$ in Eqs.~(\ref{eq:stress-stress-ff-nvt-d}) or (\ref{eq:stress-stress-ff-npt-d}).
The local stress tensor operator $\hat\sigma_{\alpha \beta}(\vec{r})$ from Eq.~(\ref{eq:definition-local-stress-tensor}) in the respective formulas must not be 
replaced by the sum of its kinetic part $[-k_{\rm\scriptscriptstyle B} T\hat\rho(\vec r)]$ plus the remaining interaction part, as it was sometimes
done in the literature (for the respective formula in the $NVT$ ensemble). Instead, the full expression in 
Eq.~(\ref{eq:definition-local-stress-tensor}) needs to be inserted in the averages in Eqs.~(\ref{eq:stress-stress-ff-nvt-d}) or (\ref{eq:stress-stress-ff-npt-d}) in order to take into account correctly the four-point momentum correlations. 

If one is
integrating out the momenta in the stress tensor operator, one can define
\begin{subequations}
\label{eq:stresses-config-space}
\begin{align}
\hat\sigma^\prime_{\alpha \beta}(\vec{r}) &= 
- k_{\rm\scriptscriptstyle B} T \hat{\rho}(\vec{r})\delta_{\alpha\beta}
+ \sum_{i<j} 
\frac{\partial\hat U}{\partial r_{ij}}
\frac{x_{ij, \alpha} x_{ij, \beta}}{r_{ij}}\,
g(\vec{r};\vec{r}_{i}, \vec{r}_j)
\label{eq:stresses-config-space-a}\\
\hat\Sigma^\prime_{\alpha \beta} &= 
- k_{\rm\scriptscriptstyle B} T \rho_\textrm{b}\delta_{\alpha\beta}
+ \frac{1}{V}\sum_{i<j} 
\frac{\partial\hat U}{\partial r_{ij}}
\frac{x_{ij, \alpha} x_{ij, \beta}}{r_{ij}}\,,
\label{eq:stresses-config-space-b}
\end{align}
\end{subequations}
where $\rho_\textrm{b}=N/V$  is the bulk density.
Note that this a fluctuating quantity in the $NPT$
ensemble.
The correlation between the local and bulk stress then becomes 
\begin{align}
	\frac{V}{k_{\rm\scriptscriptstyle B} T}
	\langle \Delta \hat\sigma_{\alpha \beta}(\vec{r}) \Delta \hat\Sigma_{\mu \nu} \rangle = &\, k_{\rm\scriptscriptstyle B} T \langle \hat{\rho}(\vec{r}) \rangle
	(\delta_{\alpha\mu} \delta_{\beta\nu} + \delta_{\alpha\nu} \delta_{\beta\mu}) \nonumber \\
	&+ \frac{V}{k_{\rm\scriptscriptstyle B} T}
	\langle \Delta \hat\sigma^\prime_{\alpha \beta}(\vec{r}) \Delta \hat\Sigma^\prime_{\mu \nu} \rangle\,.
\end{align}
This can be used in Eq.~(\ref{eq:stress-stress-ff-nvt-d}), or in Eq.~(\ref{eq:stress-stress-ff-npt-d}) if replacing $V$ by $\langle V\rangle_{\scriptscriptstyle P}$. In the correlator $\langle 
\Delta \hat{\sigma}_{\alpha \beta}(\vec{r}) \Delta V 
\big\rangle_{\scriptscriptstyle P}$ appearing in Eq.~(\ref{eq:stress-stress-ff-npt-d}) one can replace
$\hat\sigma_{\alpha\beta}(\vec{r})$ by $\hat\sigma'_{\alpha\beta}(\vec{r})$ from 
Eq.~(\ref{eq:stresses-config-space-a}).

\section{Validation for the Lennard-Jones fcc solid}
\label{sec:nnlj-application}

To validate the stress-stress fluctuation formula (\ref{eq:stress-stress-ff-npt})
in the $NPT$ ensemble, we consider the nearest-neighbor Lennard-Jones fcc solid. This model
has often been used in the literature to determine elastic constants \cite{ParrinelloRahman1981,Cowley1983,SprikImpeyKlein1984,RayMoodyRahman1985,Gusev/etal:1996,VanWorkumDePablo2003},
and thus evolved to a kind of standard test case. The pair potential between two nearest neighbors
in this model reads
\begin{equation}
\hat u(\vec{r}_{ij})= 4 \varepsilon_{\scriptscriptstyle\rm LJ} \, 
\Bigg[ \bigg( \frac{\sigma_{\scriptscriptstyle\rm LJ} }{r_{ij}} \bigg)^{12} - 
\bigg(\frac{\sigma_{\scriptscriptstyle\rm LJ} }{r_{ij}} \bigg)^6 \Bigg] \, .
\end{equation}
We use $\varepsilon_{\scriptscriptstyle\rm LJ}$ as the energy unit 
and $\sigma_{\scriptscriptstyle\rm LJ} $ as the length unit.  Hence, 
temperatures are given in units of $\varepsilon_{\scriptscriptstyle\rm LJ} /k_{\rm\scriptscriptstyle B}$, pressures and
elastic constants in units of $\varepsilon_{\scriptscriptstyle\rm LJ} /\sigma_{\scriptscriptstyle\rm LJ} ^3$, and volumes in units of 
$\sigma_{\scriptscriptstyle\rm LJ}^3$. 

\begin{table}
	\caption{Bulk elastic constants of the nearest-neighbor Lennard-Jones fcc solid in dimensionless Lennard-Jones units for two pressures $P$ 
	calculated from Monte-Carlo simulations in the $NVT$ and $NPT$ ensembles [$N=4000$, $T=0.3$]. 	
	The values were obtained by averaging over $4.5 \times10^9$ particle configurations and
	an additional symmetry-averaging over equivalent elastic constants was performed, e.~g.\ $C_{11}=(C_{11}+C_{22}+C_{33})/3$
	[cf.\ Eq.~(\ref{eq:cmatrix})]. The numerical uncertainties
	 were estimated by subdividing the sequence of simulated Monte Carlo
	configuration into independent blocks, and analyzing the fluctuations between the block averages. 
	\label{tbl:bulk-elastic-constants}}
	\begin{ruledtabular}
		\begin{tabular}{cccc}
			$P$ & $C_{\alpha \beta}$ & $NVT$ & $NPT$ \\
			\hline
			& $C_{11}$ & $44.25 \pm 0.03$ & $44.26 \pm 0.07$ \\
			$0.0$ &$C_{12}$ & $19.50 \pm 0.03$ & $19.50 \pm 0.07$ \\
			&$C_{44}$ & $23.04 \pm 0.01$ & $23.04 \pm 0.01$ \\
			\hline
			& $C_{11}$ & $60.22 \pm 0.03$ & $60.19 \pm 0.09$ \\
			$1.4$ & $C_{12}$ & $29.65 \pm 0.03$ & $29.62 \pm 0.09$ \\
			&$C_{44}$ & $30.24 \pm 0.01$ & $30.24 \pm 0.01$
		\end{tabular}
	\end{ruledtabular}
\end{table}

All presented results are obtained from 
a system containing $N=4000$ particles, corresponding to $10^3$ cubic unit cells, with periodic boundary conditions. 
We work at $T=0.3$ in the low-temperature regime, where the particles form an fcc-lattice, and compare
results obtained in the $NPT$ ensemble for two pressures $P=0$ and $P=1.4$ with that in the $NVT$ ensemble
with the corresponding volumes $V=\langle V\rangle_{\scriptscriptstyle P=0}\cong4282$ ($\rho_\textrm{b}\cong0.934$) and  $V=\langle V\rangle_{\scriptscriptstyle P=1.4}\cong4107$ ($\rho_\textrm{b}\cong0.974$). 
The simulations were carried out using standard Monte-Carlo techniques \cite{LandauBinder2014} for the both $NVT$ and $NPT$ ensembles 
in the configuration space, i.~e., without momenta. Therefore we used Eqs.~(\ref{eq:stresses-config-space}) to calculate
the non-affine contributions in Eqs.~(\ref{eq:stress-stress-ff-nvt-d}) and (\ref{eq:stress-stress-ff-npt-d}).
In the following, our results for the elastic constants $\tilde C_{\alpha\beta\mu\nu}$
are given in the usual Voigt notation \cite{Voigt1910}, where we omit the tilde in the notation.

For the bulk elastic constants, the matrix must have the form 
\begin{equation}
\begin{bmatrix}
C_{11} & C_{12} & C_{12} & 0 & 0 & 0 \\
C_{12} & C_{11} & C_{12} & 0 & 0 & 0 \\
C_{12} & C_{12} & C_{11} & 0 & 0 & 0 \\
0 & 0 & 0 & C_{44} & 0 & 0\\
0 & 0 & 0 & 0 & C_{44} & 0\\
0 & 0 & 0 & 0 & 0 & C_{44}
\end{bmatrix}
\label{eq:cmatrix}
\end{equation}
because of the cubic symmetry of the fcc lattice \cite{Nye1985}.
the three independent elastic constants $C_{11}$, $C_ {12}$ and $C_{44}$,
calculated with Eq.~(\ref{eq:stress-stress-ff-nvt}) in the $NVT$ ensemble and 
with Eq.~(\ref{eq:stress-stress-ff-npt}) in the $NPT$ ensemble (from their respective bulk version, see note at the end of Sec.~\ref{sec:nvt}),
are shown in  Table~\ref{tbl:bulk-elastic-constants}. For both ensembles, there is perfect agreement
at both simulated pressures. The results for zero pressure are also in agreement with earlier published work 
\cite{Cowley1983, Gusev/etal:1996}. Because the crystal under the higher pressure has a stronger resistance 
against deformation changes, the corresponding values for the elastic constants
are larger than for zero pressure. 

If we ignore the ensemble transformation of the non-affine contribution to the FF and 
use Eq.~(\ref{eq:stress-stress-ff-npt-d}) without the second term on the right hand side, 
the bulk elastic constant $C_{44}=\tilde C_{2323}=\partial\sigma_{23}/\partial\tilde\epsilon_{23}$ 
remains unchanged, because it refers to the shear strain $\tilde\epsilon_{23}$, where
the second term $\propto\delta_{\mu\nu}=\delta_{23}$ in Eq.~(\ref{eq:stress-stress-ff-npt-d}) vanishes.
The agreement of the expressions for the shear moduli in the $NVT$ and $NPT$ ensembles is expected also 
from the decoupling of pure deviatoric and dilatational strains \cite{Wittmer/etal:2013a}. 
However, for the bulk elastic constants $C_{11}=\tilde C_{1111}$ and $C_{12}=\tilde C_{1122}$ referring to 
normal strains,  we obtain $C_{11}=16.50$ and $C_{12}=-8.25$ for zero pressure, i.~e.\
there is a drastic deviation compared to the values $C_{11}=44.26$ and $C_{12}=19.50$
listed in Table~\ref{tbl:bulk-elastic-constants}.

\begin{figure}[t!]
	\includegraphics[width=0.48\textwidth]{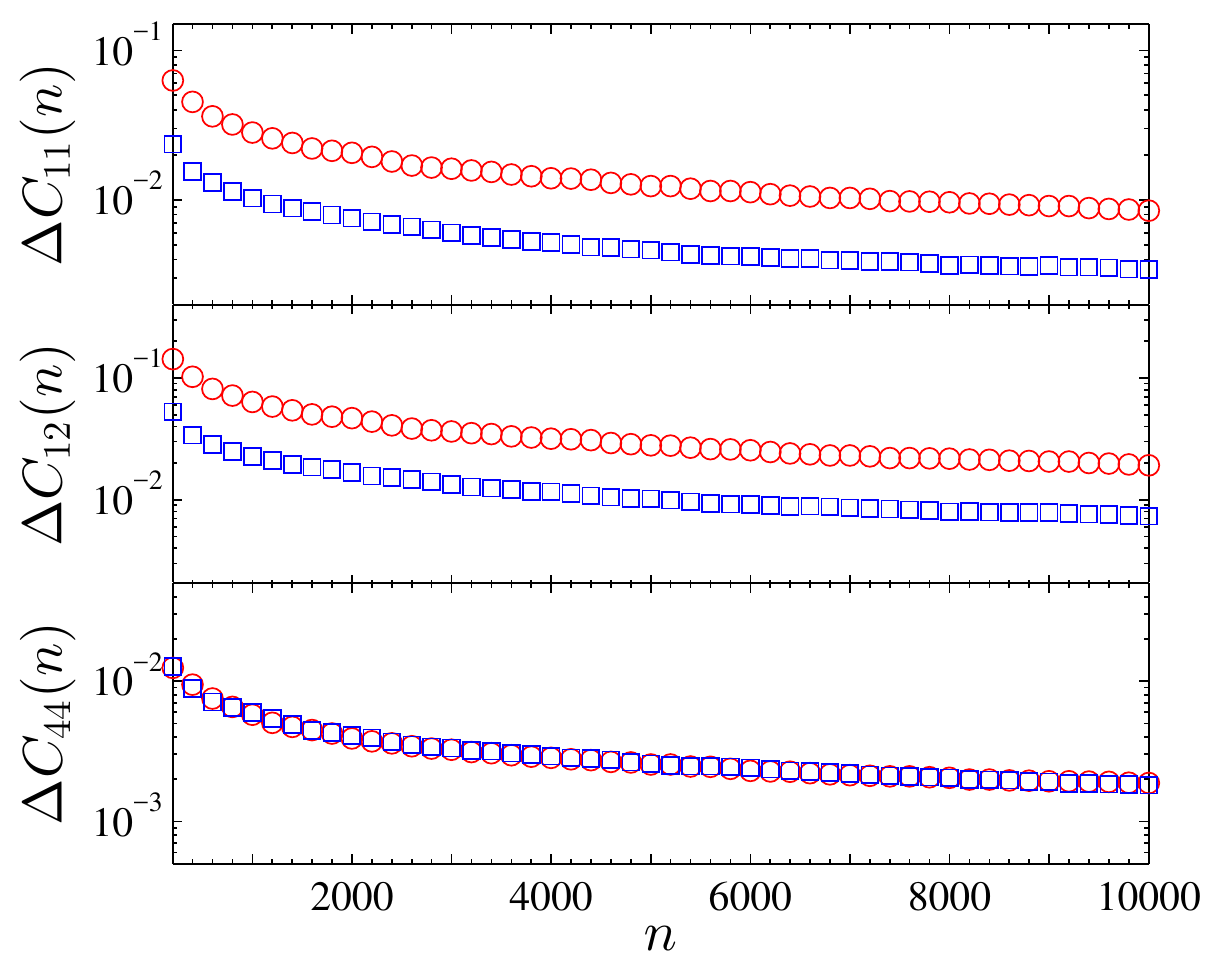}
	\caption{Mean relative deviation 
	$\Delta C_{\alpha \beta}$ 
	after averaging over $n$ independent configurations in the
	$NVT$ (blue squares) and $NPT$ ensemble (red circles) for pressure $P=0$ [$N=4000$, and $T=0.3$]. } 
 \label{fig:convergence}
\end{figure}

\begin{figure*}
\includegraphics[width=0.95\textwidth]{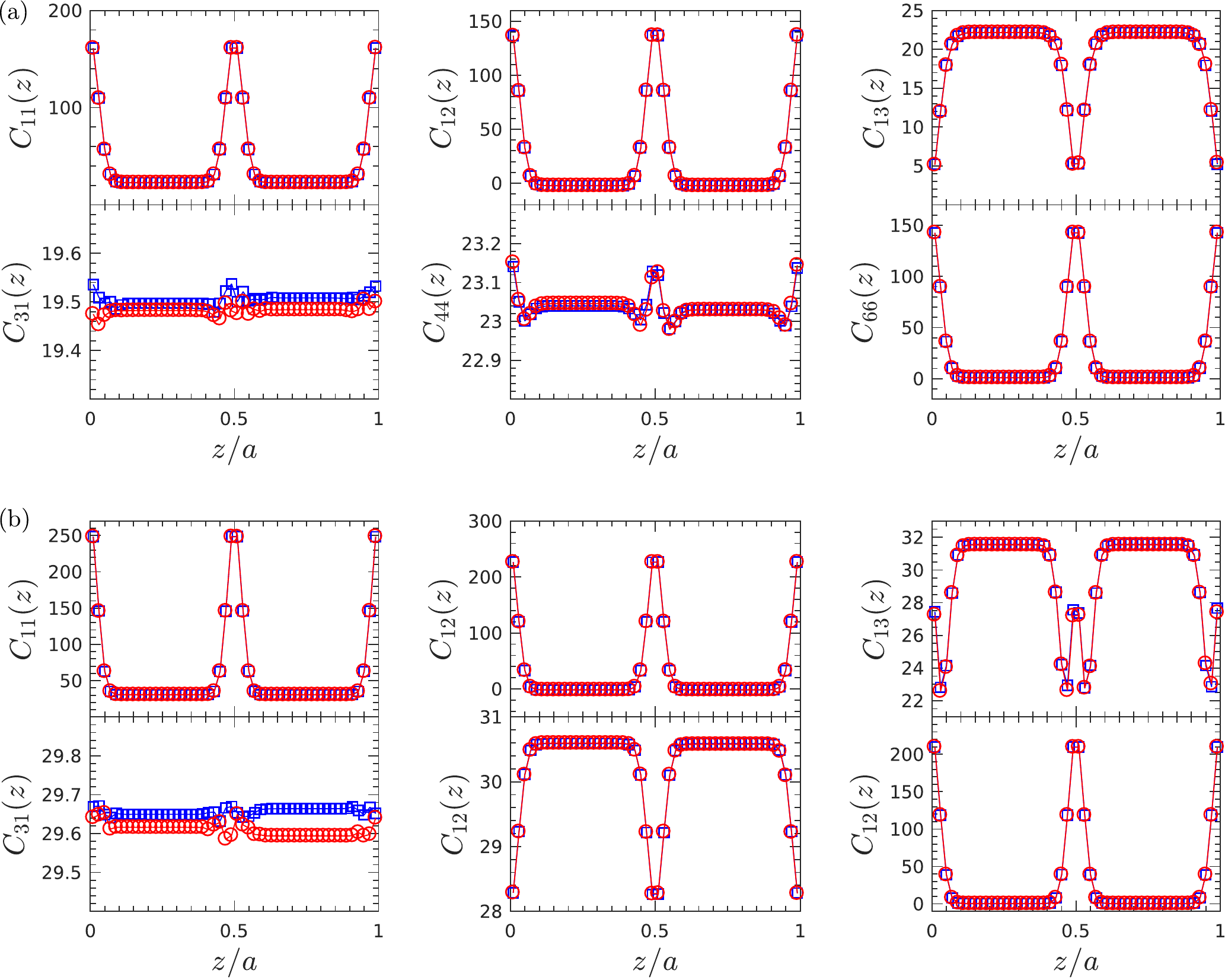}
\caption{Simulated profiles of the nonzero local elastic constants in the $NVT$ (blue squares) 
and $NPT$ ensemble (red circles) at (a) $P=0$ and (b) $P=1.4$ [$N=4000$, $T=0.3$].
The lattice constants of the cubic unit cells are $a=1.6241$ for $P=0$ and $a=1.6016$ for $P=1.4$.}
\label{fig:local-ec}
\end{figure*}

To compare the convergence behavior of the stress-stress FF in the two ensembles, we analyze the mean relative 
deviation of the bulk elastic constants with respect to their converged values in dependence of the number $n$ of configuration samples used for the ensemble averaging. 
This mean relative deviation was determined as follows. After averaging over $n$ equilibrated
configurations (separated by 20 Monte Carlo steps) in one simulation run $i$,
one obtains a value $C_{\alpha\beta}^{(i)}(n)$. This exhibits
a relative deviation $|C_{\alpha\beta}^{(i)}(n)-C^{\textrm{ref}}_{\alpha\beta}|/C^\textrm{ref}_{\alpha\beta}$,
where $C^{\textrm{ref}}_{\alpha\beta}$ refers to the converged value given in Table~\ref{tbl:bulk-elastic-constants}.
The mean relative deviation $\Delta C_{\alpha \beta} (n)$ is obtained from averaging over
a number $m=500$ of independent simulation runs $i=1,\ldots,m$, i.~e.\
$\Delta C_{\alpha \beta} (n)= (\sum_{i=1}^m 
\vert C_{\alpha\beta}^{(i)}(n) - C^\textrm{ref}_{\alpha\beta} \vert / C^\textrm{ref}_{\alpha\beta})/m$.
As can be seen from Fig.~\ref{fig:convergence}, the rate of convergence in both ensembles, i.~e.\ the decrease
of $\Delta C_{\alpha \beta} (n)$ with $n$, shows similar behavior. In the case of $C_{44}$, the mean relative
deviations are almost the same for the two ensembles, and in the case of
$C_{11}$ and $C_{12}$, they are by a factor of about two larger for the $NPT$ ensemble.
After averaging over $n=10^4$ 
samples, the mean relative deviation is of the order 1\% in both the $NVT$ and $NPT$ ensemble for $C_{11}$ and $C_{12}$, and about an order of magnitude smaller for $C_{44}$. 

Finally, we compare the variation of the local elastic constants along one principal axis of the cubic unit cell
in the two ensembles. To this end, we partition the simulation box into 500 thin slabs of equal thickness
in the $z$-direction, that means 50 slabs per length of the cubic unit cell and calculate the local elastic constants
for each slab. In the $NPT$ ensemble, where the size of the simulation box fluctuates, the slab thickness is always
adjusted accordingly. Additionally, we average over the periodicity of the crystal, which here means 
that we perform an average over the elastic constants of every 50th slab. 
Generally, one should be careful with the physical interpretation of fields of elastic constants on spatial scales
comparable to atomic distances, see the analysis and discussions in \cite{WorkumDePablo2004, Barrat:2006, Tsamados2009, Molnar/etal:2016}.
For the nearest-neighbor Lennard-Jones fcc crystal, however, it has been shown before
\cite{VanWorkumDePablo2003} that the elastic constants still relate stresses and strains linearly
according to Hooke's law even on such small scales. 

Symmetry considerations for this arrangements predict that in total 12 local elastic constants are nonzero, 
where six of them are independent. For the partitioning in $z$-direction,
these nonzero constants are $C_{11}(z)=C_{22}(z)$, $C_{12}(z)=C_{21}(z)$,
$C_{13}(z)=C_{23}(z)$, $C_{31}(z)=C_{32}(z)$, $C_{33}(z)$, $C_{44}(z)=C_{55}(z)$, and $C_{66}(z)$.
The results from the simulations confirm these predictions for both ensembles.

Figure~\ref{fig:local-ec} shows the profiles of the nonzero elastic constants $C_{\alpha\beta}(z)$ in the
$NVT$ (blue squares) and the $NPT$ ensemble (red circles) for (a) $P=0.0$ and (b) $P=1.4$. For all elastic constants
there is excellent agreement of the simulated data in the two ensembles. Due to the crystal symmetry,
the profiles are symmetric with respect to the center of the cubic unit cell with lattice constant $a$, i.~e.\
$C_{\alpha\beta}(z-a/2)=C_{\alpha\beta}(-z-a/2)$. We furthermore checked that an integration over the profiles  
gives bulk values, which exhibit the symmetries of the matrix in Eq.~(\ref{eq:cmatrix}) and agree with
the values listed in Table~\ref{tbl:bulk-elastic-constants}. 
As for the bulk values, the local elastic constants
are larger for the higher pressure. The profiles $C_{13}(z)$ and $C_{44}(z)$ change their shape with the pressure change,
where $C_{13}(z)$ shows an additional local maximum at $z=a/2$ at the higher pressure, and 
$C_{44}(z)$ exhibits a shallow local maximum at $z=a/2$ for $P=0$ and a local minimum at $P=1.4$. 

\section{Application to a lipid bilayer model}
\label{sec:lipid-application}
As an example for the application of the stress-stress FF in the $NPT$ ensemble,
we calculate the elastic constants for a simple coarse-grained model of a lipid bilayer
as developed by Lenz and Schmid (LS model) \cite{Lenz/Schmid:2005, Lenz/Schmid:2007, Lenz:2007}. 
This model is a coarse-grained representation of single-tail amphiphilic molecules, where the
hydrophilic part is represented by one head bead and the long aliphatic tail by six tails
beads. Adjacent beads in one molecule are connected by finitely
extensible nonlinear elastic (FENE) springs \cite{Grest/Kremer:1986}
with potential
\begin{equation}
V_{\scriptscriptstyle\rm FENE}(r)=-\frac{v_{\scriptscriptstyle\rm FENE}}{2}
(\Delta r_\textrm{m})^2
\log\left[1-\left(\frac{r-r_0}{\Delta r_\textrm{m}}\right)^2\right]\,,
\label{eq:vfene}
\end{equation}
where $v_{\scriptscriptstyle\rm FENE}$ specifies the strength of the
springs and $r$ the bead distance; $r_0$ is the bond length for the unstretched spring
and $\Delta r_\textrm{m}$ is the maximal stretching distance. Three
adjacent beads of a molecule with bond angle $\theta$ 
interact via the  bond-angle potential
\begin{equation}
V_{\rm ba}(\theta)=v_{\rm ba}\left(1-\cos\theta\right)\,,
\label{eq:vba}
\end{equation}
where $v_{\rm ba}$ regulates the stiffness of the chain molecules.
Both the non-bonded beads belonging to the same molecule and the beads
belonging to different molecules interact via a truncated
Lennard-Jones potential
\begin{subequations}
\begin{align}
V_{\rm sc}(r)&=\left[V_{\scriptscriptstyle \rm LJ}(r)-V_{\scriptscriptstyle \rm \rm LJ}(r_\textrm{c})\right]
H(r_\textrm{c}-r)\,,\label{eq:vsc-a}\\[1ex]
V_{\scriptscriptstyle\rm LJ}(r)&=v_{\scriptscriptstyle\rm LJ}
\left[\left(\frac{\sigma_{\rm\scriptscriptstyle LJ}}{r}\right)^{12}
-2\left(\frac{\sigma_{\rm\scriptscriptstyle LJ}}{r}\right)^6\right]\label{eq:vsc-b}\,,
\end{align}
\label{eq:vsc}
\end{subequations}
where $H(.)$ is the Heaviside step function [$H(x)=1$ for
$x>0$ and zero otherwise]. The parameter
$v_{\scriptscriptstyle\rm LJ}$ is the same for all types of the interacting beads and used
as the energy unit. The $\sigma_{\rm\scriptscriptstyle LJ}$
are different for different types of beads, see Table~\ref{tab:parameters}. The length unit is set by
$\sigma_{\rm\scriptscriptstyle LJ}$ for the tail-tail bead interactions.
These units correspond to
$v_{\scriptscriptstyle\rm LJ}\simeq0.36\times10^{-20}$~J and
$\sigma_{\scriptscriptstyle\rm LJ}\simeq6$~\AA\
\cite{Lenz/Schmid:2005,Lenz:2007}. 
The cutoff parameters $r_\textrm{c}$ correspond to the
minimum of $V_{\scriptscriptstyle\rm LJ}(r)$ for head-head and head-tail bead
interactions (i.~e., $r_\textrm{c}=\sigma_{\scriptscriptstyle\rm LJ}$ in that case), leading to
a purely repulsive interaction between these types of beads. For the tail-tail bead interactions, 
$r_\textrm{c}$ has a larger value, giving an attractive pair interaction for
$\sigma_{\scriptscriptstyle\rm LJ}<r<r_\textrm{c}$. This attractive part facilitates a self-organization 
of the chain molecules into a double-layer structure.

To stabilize a lipid bilayer structure, the bead-spring chains representing the molecules
are brought into contact with a solvent, which is also represented by simple beads. These
interact
with all lipid beads via $V_{\rm sc}(r)$ with parameters given in Table~\ref{tab:parameters} (solely
repulsive interactions),
but they do not interact with themselves.
All parameters of the LS model are
summarized in Table~\ref{tab:parameters}. Units 
of temperature and pressure (as well as the elastic constants) 
are $v_{\scriptscriptstyle\rm LJ}/k_{\rm\scriptscriptstyle B}$ and 
$v_{\scriptscriptstyle\rm LJ}/\sigma_{\scriptscriptstyle\rm LJ}^3$,
with $\sigma_{\scriptscriptstyle\rm LJ}$ for tail-tail bead interactions.
As mentioned above in Sec.~\ref{sec:nvt}, the truncation of the Lennard-Jones at the radii $r_\textrm{c}$ in Table~\ref{tab:parameters}
gives rise to discontinuities in the second-order derivatives of the respective potential and accordingly 
$\delta$-function contributions in the Born term, see Eqs.~(\ref{eq:bornfunction}) and 
(\ref{eq:stress-stress-ff-npt-c}). The corresponding impulsive contributions were taken into account 
for bead with pair distances $r$ in an interval $[r_\textrm{c}-\Delta r/2,r_\textrm{c}+\Delta r/2]$ with $\Delta r=0.02$ 
\footnote{A more sophisticated method for treating the impulsive contributions is given in \cite{Xu/etal:2012}}.

\begin{table}[t!]
  \caption{Parameters of the interaction potentials 
    in Eqs.~(\ref{eq:vfene})-(\ref{eq:vsc}) \cite{Lenz:2007}, in units
    of the Lennard-Jones parameters $v_{\scriptscriptstyle\rm LJ}=1$ and
    $\sigma_{\scriptscriptstyle\rm LJ}=1$ 
    for the tail-tail interactions.}
\label{tab:parameters}
\begin{tabular}{|c|c|c|}\hline
interaction type & potential & parameters\\[1ex] \hline\hline
tail-tail & & $v_{\scriptscriptstyle\rm LJ}=1$, 
    $\sigma_{\scriptscriptstyle\rm LJ}=1$, $r_\textrm{c}=2$\\[1ex] \cline{1-1}\cline{3-3}
head-tail & & \\ 
solvent-tail & \raisebox{0ex}[0ex]{$V_{\rm sc}$}
    & \raisebox{1.5ex}[-1.5ex]{$v_{\scriptscriptstyle\rm LJ}=1$, 
    $\sigma_{\scriptscriptstyle\rm LJ}=1.05$, $r_\textrm{c}=1.05$}\\[1ex] \cline{1-1}\cline{3-3}
head-head & & \\
solvent-head & & \raisebox{1.5ex}[-1.5ex]{$v_{\scriptscriptstyle\rm LJ}=1$, 
    $\sigma_{\scriptscriptstyle\rm LJ}=1.1$, $r_\textrm{c}=1.1$}\\[1ex] \hline
solvent-solvent & none & \\[1ex] \hline
bond length & $V_{\scriptscriptstyle\rm FENE} $ & $v_{\scriptscriptstyle\rm FENE}=100$, 
        $r_0=0.7$, $\Delta r_m=0.2$\\[1ex] \hline
bond angle & $V_{\rm ba}$ & $v_{\rm ba}=4.7$\\[1ex] \hline
\end{tabular}
\end{table}

Monte-Carlo simulations were performed as described in
\cite{Lenz/Schmid:2005,Lenz:2007} under constant temperature $T=1.3$ and pressure $P=2$ 
for lipid bilayers with upper and lower leaflets consisting of
$N=10\times10$ lipid molecules; 7692 solvent beads were chosen for the solvent model. 
The lipid bilayer is oriented in the $xy$-plane and the center of mass of the lipid beads
defines the origin of the coordinate system.

\begin{figure}[t!]
	\includegraphics[width=1.0\columnwidth]{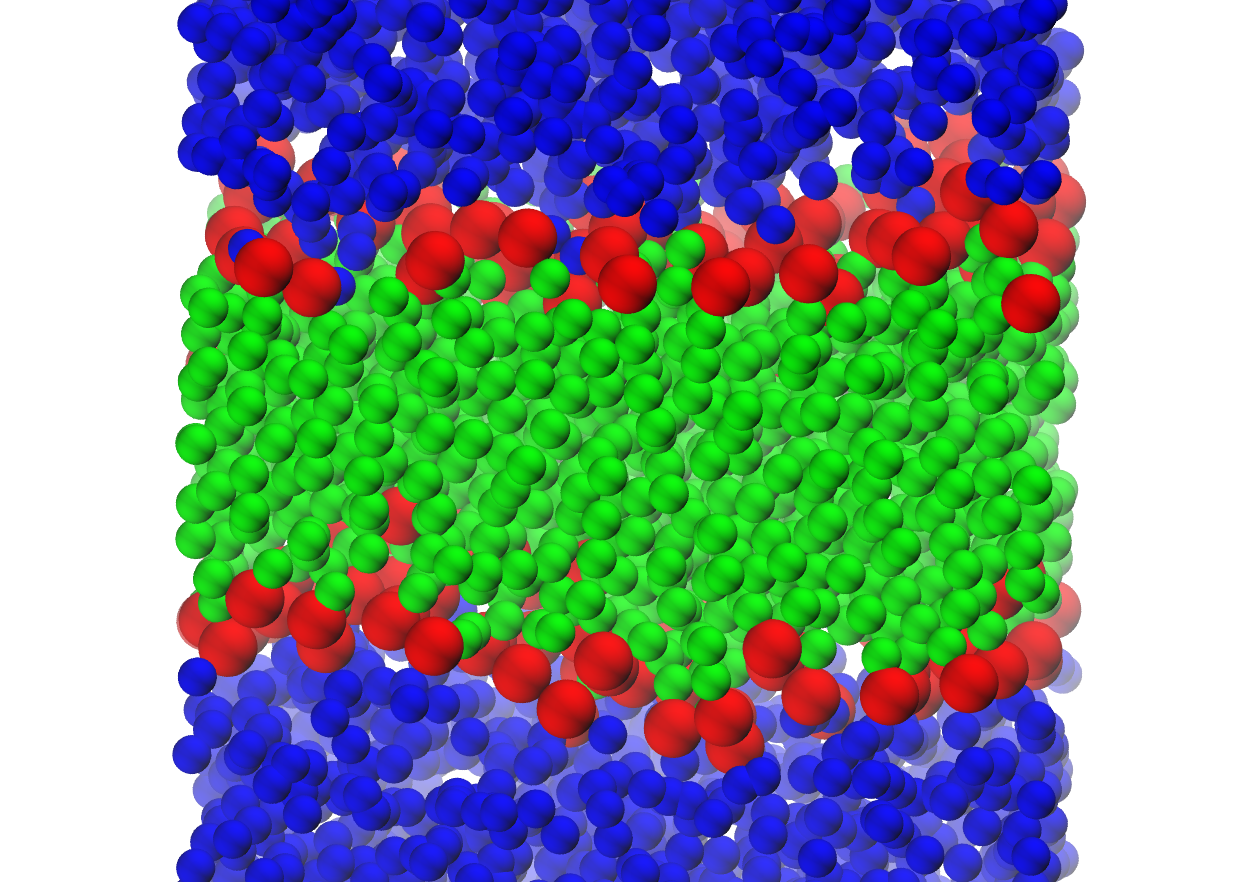}
	\caption{Snapshot of an equilibrated particle configuration of a lipid bilayer in the fluid phase
	together with the surrounding 
	solvent from Monte-Carlo simulations of the LS model. Solvent beads are colored in blue, head beads
	in red, and tail beads in green.}
	\label{fig:ls-configuration}
\end{figure}

A representative example of an equilibrated configuration of the system with lipid bilayer and
solvent beads is shown in Fig.~\ref{fig:ls-configuration}.
Because of the rotational symmetry around the $z$-axis,
the elastic constants of the full lipid bilayer should exhibit transverse symmetry,
corresponding to the block diagonal form 
\begin{equation}
\begin{bmatrix}
C_{11} & C_{12} & C_{13} & 0 & 0 & 0 \\
C_{12} & C_{11} & C_{13} & 0 & 0 & 0 \\
C_{13} & C_{13} & C_{33} & 0 & 0 & 0 \\
0 & 0 & 0 & C_{44} & 0 & 0\\
0 & 0 & 0 & 0 & C_{44} & 0\\
0 & 0 & 0 & 0 & 0 & C_{66}
\end{bmatrix}
\label{eq:cmatrix-lipid}
\end{equation}
of the tensor in Voigt notation.

\begin{figure}[t!]
\includegraphics[width=1.0\columnwidth]{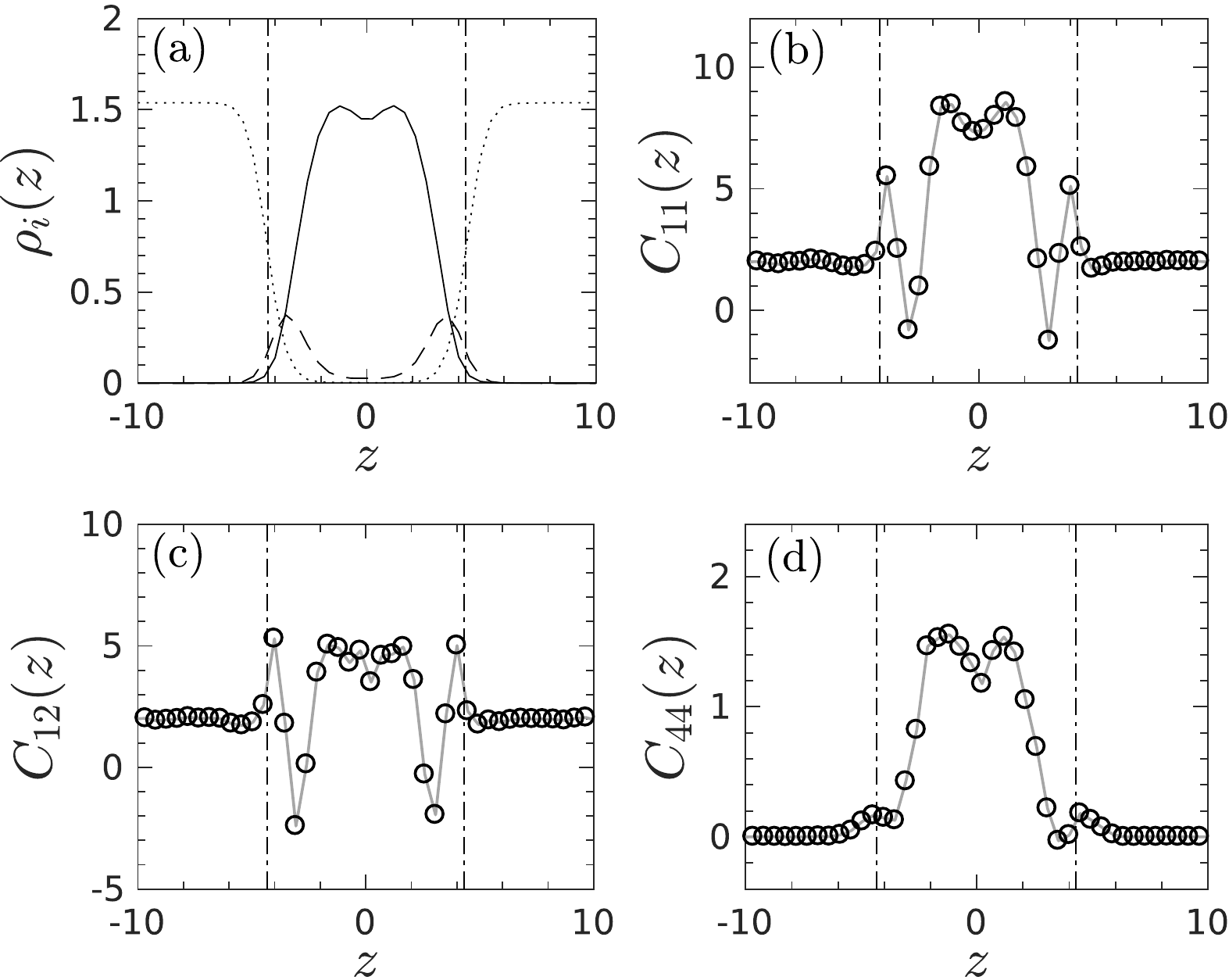}
\caption{(a) Density profiles $\rho(z)$ of head beads (dashed line), 
tail beads (solid line) and solvent beads (dotted line).
(b)-(d) Representative profiles $C_{\alpha\beta}(z)$ of 
elastic constants. The vertical lines mark the boundaries of the lipid bilayer region
[see text and Eq.~(\ref{eq:c-full})]. Results were obtained from Monte-Carlo simulations
at $T=1.3$ and $P=2$.}
\label{fig:local-ec-lipid}
\end{figure}

To determine the $C_{\alpha\beta}$ of the full lipid bilayer, we can now take advantage of
the access to local elastic constants, which allows us to selectively average them
over the region of the bilayer. Technically, we 
calculate the local constants with respect to the $z$-coordinate
using Eqs.~(\ref{eq:stress-stress-ff-npt})
by dividing the simulation
box in 100 slabs with respect to the (instantaneous) box length in $z$-direction, 
i.~e., we use the same method as described above in Sec.~\ref{sec:nnlj-application}.
The average slab thickness was $\Delta z=0.47$. The three-body contributions to 
the stress tensor and the Born-term resulting from the bond-angle potential in Eq.~(\ref{eq:vba})
are decomposed into pairwise contributions according to \cite{VanWorkum2006}. 
In Eq.~(\ref{eq:stress-stress-ff-npt-d}) 
we take the ideal gas value $K=\rho_\textrm{s}k_{\rm\scriptscriptstyle B}T=P$, with $\rho_\textrm{s}$
the constant density of the solvent beads far from the lipid bilayer, see Fig.~\ref{fig:local-ec-lipid}(a)
\footnote{In the thermodynamic limit, the lipid beads would occupy a zero fraction of the total volume,
yielding a bulk modulus $K=\rho_\textrm{s}k_{\rm\scriptscriptstyle B}T$ for the whole system.
For a small system, the bulk modulus shows deviations from its thermodynamic limit due to 
contributions from the lipid bilayer.}.

Examples of the resulting profiles $C_{\alpha\beta}(z)$
are shown in Fig.~\ref{fig:local-ec-lipid}(b)-(d), together with the density profiles of the
head beads (dashed line), tail beads (solid line), and solvent beads (dotted line) in 
Fig.~\ref{fig:local-ec-lipid}(a). The two profiles in Figs.~\ref{fig:local-ec-lipid}(b) and (c) are 
representative of $C_{\alpha\beta}$ belonging to the first block diagonal element
of (\ref{eq:cmatrix-lipid}) [$\alpha,\beta\in\{1,2,3\}$]: They are positive in the tail
bead region and show an oscillation near the lipid-solvent interface, where
the head beads have noticeable density, see Fig.~\ref{fig:local-ec-lipid}(a).
Near the interface a small $z$-interval exists, where $C_{11}$ and $C_{12}$
become negative.
The profile in Fig.~\ref{fig:local-ec-lipid}(d) is
representative of the nonzero $C_{\alpha\beta}$ belonging to the second block diagonal element
of (\ref{eq:cmatrix-lipid}) [$\alpha=\beta\in\{4,5,6\}$]. Here the impact of the head beads
at the lipid-solvent interface is much less pronounced.
All profiles in Figs.~\ref{fig:local-ec-lipid}(b)-(d) exhibit a small dip
close to the midplane of the bilayer at $z=0$ ("leaflet-interface") and
reflect well the spatial symmetry with respect to this midplane.

In the present case, the zero elements in (\ref{eq:cmatrix-lipid}) 
should give zero values also on the local scale. Indeed,
we found the profiles calculated from the simulations to fluctuate around zero with a standard deviation of
0.42. In the solvent region, all profiles in Figs.~\ref{fig:local-ec-lipid}(b)-(d) are flat 
with value $k_{\rm\scriptscriptstyle B}T\rho(z)=K=P=2$ for the $C_{\alpha\beta}(z)$, $1\le\alpha,\beta\le3$,
and zero value for the other $C_{\alpha\beta}(z)$. This is the expected behavior
for the mutually non-interacting solvent beads, which correspond to an ideal gas.
Since we are interested in the elastic constants of the full lipid bilayer, we are not
investigating further here, on which spatial scale the profiles in Figs.~\ref{fig:local-ec-lipid}(b)-(d)
reflect a linear relation between local stresses and strains according to Hooke's law.

By a spatial averaging over the profiles $C_{\alpha\beta}(z)$
the elastic constants of the full lipid bilayer are obtained,
\begin{equation}
	C^{\scriptscriptstyle \textrm{LB}}_{\alpha\beta} =  \frac{1}{(d_+-d_-)}
	\int_{d_-}^{d_+} \hspace{-0.5em}
	\mathrm{d} z\, C_{\alpha\beta}(z) \, ,
\label{eq:c-full}
\end{equation}
where $d_\pm=z_\pm\pm\sigma_{\scriptscriptstyle\rm LJ}$ with
$\sigma_{\scriptscriptstyle\rm LJ}$ the value for head-solvent interactions (see Table~\ref{tab:parameters}), 
and $z_-$ and $z_+$
are the average $z$-coordinates of the head bead in the lower and upper leaflet, respectively.
By shifting these position with  $\sigma_{\scriptscriptstyle\rm LJ}$
we take into account the soft core interaction range between head and solvent
beads. The positions $d_-=-4.33$ and $d_+=4.30$ are marked by the vertical lines in 
Figs.~\ref{fig:local-ec-lipid}(a)-(d). As can be seen from Fig.~\ref{fig:local-ec-lipid}(a), they
represent well the lower and upper limit of the bilayer. In Figs.~\ref{fig:local-ec-lipid}(b)-(d)
we see that they mark also the points, where the profiles of the local elastic constants cross over
to the flat solvent regime.

The tensor of elastic constants of the full lipid bilayer obtained from Eq.~(\ref{eq:c-full}) is
\begin{equation}
C^{\scriptscriptstyle \textrm{LB}}\!=\! 
\left[\begin{array}{rrrrrr}
5.11 & 3.00 & 2.80 & -0.14 & -0.22 & -0.02 \\ 
3.03 & 5.34 & 3.08 & -0.04 & 0.01 & -0.03 \\ 
2.67 & 2.94 & 4.15 & -0.06 & -0.34 & 0.02 \\ 
-0.15 & -0.04 & -0.05 & 0.91 & 0.02 & -0.02 \\ 
-0.21 & 0.01 & -0.33 & 0.02 & 0.70 & -0.04 \\ 
-0.03 & -0.03 & 0.03 & -0.03 & -0.04 &1.18
\end{array}\right]\,.
\label{eq:cmatrix-result}
\end{equation}
We see that the large elements are
the $C_{\alpha\beta}$ with $1\le\alpha,\beta\le3$, while values an order of magnitude smaller are
obtained for those matrix elements, which should be zero according to the 
expected form in (\ref{eq:cmatrix-lipid}). From these elements we can estimate that the numerical
uncertainty is about $\pm0.3$. In terms of this numerical uncertainty, the diagonal elements $C_{44}$ and $C_{66}$, despite being small, have significant values, and the structure of (\ref{eq:cmatrix-result}) 
agrees with the expected symmetry of (\ref{eq:cmatrix-lipid}).

\section{Conclusions}
\label{sec:conclusion}

We derived a new stress-stress FF to calculate the tensors of bulk and local elastic constants
in the $NPT$ ensemble based on formerly derived expressions for the $NVT$ ensemble and transformation
rules for correlation functions between different ensembles. This FF allows the determination of elastic constants
from simulations in the $NPT$ ensemble. We validated the FF for the
nearest-neighbor Lennard-Jones solid and showed the agreement of the results with those obtained
from corresponding simulations in the $NVT$ ensemble.
As an application, we calculated the tensor of elastic constants for a simulated lipid bilayer in the fluid phase.

For solid materials, the new FF for the $NPT$ ensemble can facilitate
the analysis of the pressure dependence of elastic constants.
An efficient procedure to calculate elastic
constants in the $NPT$ ensemble should in particular be useful for systems, which naturally
need to be held under an external pressure.
These are often soft matter structures forming in aqueous environments and
exhibiting an elastic response behavior for small deformations, for example, 
lipid membranes or cell organelles.
The access to local elastic constants provides a means for detailed analyses of heterogeneous systems,
for which the studied lipid bilayer gives an example. As was shown recently, it is possible
also to extend the methodology for studying time-dependent relaxation behavior of elastic moduli
\cite{Wittmer/etal:2015}.

Other interesting systems are
structured composite materials, as, for example, sandwich materials
with layer structure. A typical approach to calculate the elastic properties of such 
composite is to perform a weighted average
over elastic properties of a homogeneous material assigned to each layer.
Below a certain layer thickness, where the properties inside a layer become strongly influenced by
interfacial effects, such calculation based on a simple layer representation
will break down and the FF could then be used.
From the theoretical point of view, one can study such composite systems with
particle-based models and systematically analyze
how large the substructures must be that a calculation based on a
simple substructure representation becomes valid.
Here it would be interesting to see whether one can find simple approximate rules,
e.~g.\ with respect to the influence of the particle interaction range.

Self-organized structures are often formed by complex molecules with interactions
involving many-body forces, and these occur also for atomic systems when using 
accurate effective potentials derived from first-principle calculations.  
In the bulk, the stress-stress FF in the $NVT$ ensemble can be formulated for
arbitrary many-body forces \cite{Lutsko:1989} and based on this, it is possible to
take over the formulas in the $NPT$ ensemble as described. However, in the presence of 
many-body forces, it becomes more difficult to define a local stress tensor and 
associated stress-stress FF for local elastic constants.
In some cases,  as, e. g., for the bond-angle potential in Eq.~(\ref{eq:vba}), it is possible
to decompose many-body forces into pair forces \cite{VanegasTorresArroyo2014, Torres2015, 
VanWorkum2006}. 
For the general case, it would be desirable to investigate in more detail how
many-body forces can be effectively handled in the determination
of elastic constants.

%\bibliography{elasticity-min}


\end{document}